\documentstyle[preprint,tighten,aps,psfig]{revtex}

\newcommand{\real}{{\sf I}\kern-.12em{\sf R}}

\begin{document}
\draft
\preprint{IFUP-TH 35/98, UCY-PHY 8/98}
\title{
Resummation of Cactus Diagrams in the Clover Improved 
Lattice Formulation of QCD}
\author{H. Panagopoulos$^a$, E. Vicari$^b$}
\address{
$^a$Department of Natural Sciences, University of Cyprus,
P.O.Box 537, Nicosia CY-1678, Cyprus. haris{@}ucy.ac.cy}
\address{
$^b$Dipartimento di Fisica dell'Universit\`a 
and I.N.F.N., 
via Buonarroti 2, I-56126 Pisa, Italy. vicari{@}ibmth.difi.unipi.it}

%\date{\today}

\maketitle

\begin{abstract}
We extend to the clover improved lattice formulation of QCD
the resummation of cactus diagrams, i.e. a certain
class of tadpole-like gauge invariant diagrams.
Cactus resummation yields an improved perturbative expansion.
We apply it to the lattice renormalization of some two-fermion operators
improving their one-loop perturbative estimates.


\end{abstract}

%\newpage

% ========================= BODY =========================
%\narrowtext

\bigskip
\bigskip

In a previous work \cite{P-V} we showed how to perform a resummation 
of a certain class of gauge invariant diagrams, termed cactus diagrams,
in the Wilson formulation (for both gluons and fermions) of
lattice QCD. The  resummation of such diagrams led to an 
improved perturbative expansion, essentially by dressing the one-loop
calculation of the lattice renormalizations.
Applied to a number of cases of interest, this expansion 
yielded a remarkable improvement when compared with the available
nonperturbative estimates.
In this paper we extend such calculations to the case of the clover improved
action formulation of lattice QCD~\cite{S-W}, which is widely used in numerical
simulations  in order to reduce scaling corrections.
In the following we will heavily refer to Ref.~\cite{P-V} for 
notation and many analytical results.

\noindent
\begin{minipage}{0.68\linewidth}
\ \ \ \ \ \ 
Cactus diagrams 
are tadpole diagrams which become disconnected if any one of their
vertices is removed (see Fig.~1). Our original motivation was the well known
observation of ``tadpole dominance'' in lattice perturbation theory.
Indeed tadpoles diagrams are often largely responsible for lattice artifacts.
This observation has already inspired many proposals to improve lattice pertubation
theory, see e.g. \cite{Parisi,L-M}.
Of course the contribution of standard tadpole diagrams is not gauge
invariant. So we need to further specify the class of gauge invariant
diagrams we are considering.

\end{minipage}\hskip0.02\textwidth
\begin{minipage}{0.30\linewidth}
\medskip
\hskip1.0cm\psfig{figure=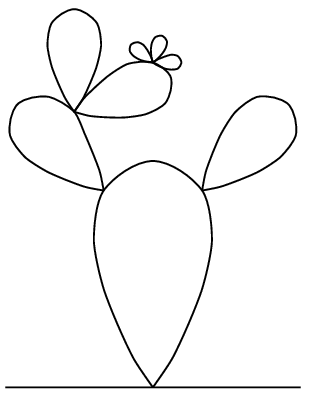,width=3truecm}\hskip1.0cm
\bigskip
{\centerline{\bf Figure 1: A cactus}}
\end{minipage}

Let us write the so-called clover improved action
\begin{eqnarray}
S_L = && {1\over g_0^2} \sum_{x,\mu\nu}
{\rm Tr}\left[ 1 - U_{x,\mu\nu} \right]  +
\sum_{f}\sum_{x} (4+m_{f,0})\bar{\psi}^{f}_x\psi^f_x 
\nonumber \\
&& -{1\over 2}\sum_{f}\sum_{x,\mu}
\left[ 
\bar{\psi}^f_x\left( 1 - \gamma_\mu\right)
U_{x,\mu}\psi^f_{x+\hat{\mu}} +
\bar{\psi}^f_{x+\hat{\mu}}\left( 1 + \gamma_\mu\right)
U_{x,\mu}^\dagger
\psi^f_x\right]  \nonumber \\
&& + c_{\rm{SW}} \sum_{f}\sum_{x,\mu\nu}
\bar{\psi}^f_x {i\over 4} \sigma_{\mu\nu}\widehat{F}_{x,\mu\nu}
\psi^f_x, 
\label{latact}
\end{eqnarray}
where $f$ is a flavor index;
$U_{x,\mu\nu}$ is the usual product of link variables
$U_{x,\mu}$ along the perimeter of a plaquette
originating at $x$ in the positive $\mu$-$\nu$ directions;
\begin{eqnarray}
\widehat{F}_{x,\mu\nu}&=&
{1\over 8} \left( Q_{x,\mu\nu} - Q_{x,\nu\mu}\right),\\
Q_{x,\mu\nu} &=& 
U_{x,\mu}U_{x+\hat{\mu},\nu}U_{x+\hat{\nu},\mu}^\dagger U_{x,\nu}^\dagger 
+  U_{x,\nu}U_{x-\hat{\mu}+\hat{\nu},\mu}^\dagger
U_{x-\hat{\mu},\nu}^\dagger U_{x-\hat{\mu},\mu} \nonumber \\ 
&+&
U_{x-\hat{\mu},\mu}^\dagger
U_{x-\hat{\mu}-\hat{\nu},\nu}^\dagger
U_{x-\hat{\mu}-\hat{\nu},\mu} U_{x-\hat{\nu},\nu} +
U_{x-\hat{\nu},\nu}^\dagger
U_{x-\hat{\nu},\mu}
U_{x+\hat{\mu}-\hat{\nu},\nu} U_{x,\mu}^\dagger .
\label{FFF}
\end{eqnarray}
The improvement coefficient $c_{\rm SW}$ can be calculated in perturbation
theory as a function of $g_0^2$. Its tree-order value is $c_{\rm SW}=1$;
in this case only the leading log scaling corrections of $O(a)$ are eliminated.
More recently a nonperturbative determination has also been performed,
which
allows to completely cancel the $O(a)$ corrections~\cite{L-S-S-We,L-S-S-W-W}.

By the Baker-Campbell-Hausdorff (BCH) formula we have:
\begin{eqnarray}
U_{x,\mu\nu}&& = 
e^{i g_0 A_{x,\mu}} e^{i g_0 A_{x+\mu,\nu}} e^{-i
g_0 A_{x+\nu,\mu}} e^{-i g_0 A_{x,\nu}} \nonumber \\
&&=\exp\left\{i g_0 (A_{x,\mu} + A_{x+\mu,\nu} -
A_{x+\nu,\mu} - A_{x,\nu}) + {\cal O}(g_0^2) \right\} \nonumber \\
&&=\exp\left\{ i g_0 F_{x,\mu\nu}^{(1)} +  i g_0^2
F_{x,\mu\nu}^{(2)} + {\cal O}(g_0^4)
\right\}
\end{eqnarray}
The diagrams that we propose to resum to all orders are the cactus
diagrams made of vertices containing $F_{x,\mu\nu}^{(1)}\,$.
Terms of this type come from the pure gluon  and clover parts of
the lattice action.

In Ref.~\cite{P-V} we showed how these diagrams dress the gluon
propagator and the gluon vertices
(we denote by a thick (thin) solid line the
transverse dressed (bare) gluon propagator):
\begin{equation}
\psfig{figure=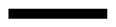,width=1truecm} = 
\psfig{figure=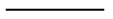,width=1truecm} \cdot 
{1\over 1-w(g_0)}
\label{propdr}
\end{equation}
where the function $w(g_0)$ can be extracted by an
appropriate algebraic equation that
 has been derived in Ref.~\cite{P-V} and that can be easily
solved numerically; for $SU(3)$, $w(g_0)$ satisfies:
\begin{equation}
u \, e^{-u/3} \, \left[u^2 /3 - 4u +8\right]  = 2 g_0^2, \qquad 
u(g_0) \equiv {g_0^2 \over 4 (1-w(g_0))}.
\end{equation}
The 3-point vertex dresses as:
\begin{equation}
\psfig{figure=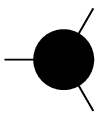,width=1truecm} \ \ {= \atop \phantom{0}}  \ \ 
\psfig{figure=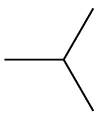,width=1truecm}\  {{\displaystyle \cdot \ 
(1-w(g_0)) }\atop 
\phantom{0}} 
\label{THREEPOINT}
\end{equation}
and similarly for other vertices.
Contributions to vertices coming from the standard Wilson
fermionic action stay unchanged, since their definition contains no
plaquettes on which to apply the linear BCH formula. 
In the clover improved action formulation  plaquettes appear 
in the new fermionic term; thus in this case one should 
also dress the new fermion-gluon vertices originating from
this term. 

Let us now prove that the fermion-gluon three-point vertex coming from
the clover term gets dressed as the three-gluon vertex,
cf. Eq.~(\ref{THREEPOINT}).
Proceeding as in Ref.~\cite{P-V} (cf. Eq. (20) and App. B therein), we write for the
fermion-gluon three-point vertex:
\begin{eqnarray}
\psfig{figure=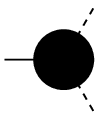,width=1truecm} \ \  
&&{\displaystyle = \atop \phantom{0}} \ \ 
\psfig{figure=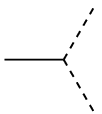,width=1truecm}\ \  {\displaystyle + \atop \phantom{0}} \ \ 
\psfig{figure=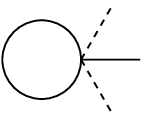,width=1truecm}\ \  {{\displaystyle \cdot 
\  {1\over 1-w(g_0)}\ \   + }\atop \phantom{0}}
\ \ 
\psfig{figure=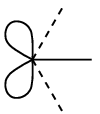,width=1truecm} {{\displaystyle \cdot \ {1\over
[1-w(g_0)]^2} + \cdots} \atop \phantom{0}}\nonumber \\
&&{\displaystyle = \atop \phantom{0}} \ \ 
\psfig{figure=gfbare.eps,width=1truecm}\ \ 
{{\displaystyle \cdot \left\{ \sum_{j=0}^\infty  {(i g_0)^{2j} \over (2j+1)!} \cdot 
{2 \, F(2j+2;N)\over N^2 {-}1} \cdot \left({1\over 2}\right)^j \cdot \ {1\over
[1-w(g_0)]^j} \right\}} \atop \phantom{0}}\nonumber \\
&&{\displaystyle = \atop \phantom{0}} \ \ 
\psfig{figure=gfbare.eps,width=1truecm}\ \
{{\displaystyle \cdot \, [1-w(g_0)]} \atop \phantom{0}}
\label{gfdressed}
\end{eqnarray}
(solid (dashed) lines represent gluons (fermions)).
No other dressed vertices are necessary in most of the interesting applications,
that essentially amount to a dressing of the perturbative one-loop calculation.
In these cases the dressing of the fermion-gluon three-point
vertex in the one-loop calculation  is equivalent to a rescaling of the
constant $c_{\rm{SW}}$:
\begin{equation}
c_{\rm{SW}} \longrightarrow \bar{c}_{\rm{SW}}\equiv c_{\rm{SW}}
\cdot \left( 1 - w(g_0)\right).
\label{cswresc}
\end{equation}

One can apply 
the resummation of cactus diagrams to the calculation of
the renormalizations of lattice operators.
Approximate expressions are
obtained by dressing the corresponding one-loop
calculations.
In the case of operators whose anomalous
dimension is zero in the $\overline{\rm MS}$ renormalization scheme,  
a consistent means of implementing the cactus dressing is to apply
it to the one-loop difference between lattice and continuum
contributions that determine the finite renormalization.
Cases with nonzero anomalous dimension can be dealt
with in an analogous manner,
by setting the scale $\mu=1/a$ and dressing the finite renormalization coefficients 
as before.

In the following we present a few examples of lattice renormalizations
for which
non-pertubative evaluations are available
in the literature. 
Let us consider the 
non-singlet vector and axial
currents $V_\mu^a=\bar{\psi} \lambda^a\gamma_\mu \psi$ and
$A_\mu^a=\bar{\psi} \lambda^a\gamma_\mu \gamma_5\psi$
and the renormalization of their lattice counterparts.
So far, essentially three nonperturbative methods have been 
successfully implemented in the computation of such renormalizations:
(i) use of the Ward identities~\cite{romani} (WI) ;
(ii) nonperturbative renormalization on external quark and gluon states~\cite{M-P-S-T-V}
(NP);
(iii) use of finite size scaling techniques~\cite{L-S-S-W} (FSS).
The major source of systematic error in these calculations is due
to $O(a)$ scaling violations. Already for the
tree clover improved action they  turn out to be rather small
at $g_0^2\simeq 1$,  where simulations are actually done.
So nonperturbative estimates are quite reliable. 
In the case of the tree clover improved action, scaling
corrections are estimated to be less than 5\% at $g_0^2\simeq 1$
using the WI approach~\cite{M-S-V,C-L-V}.
Lattice renormalizations can be also calculated in perturbation theory.
Most perturbative calculations have been performed to one loop.
Thus their use as approximation of the lattice renormalizations
introduces $O(g_0^4)$ errors in the final estimates of physical quantities
(to be compared with
the $O(a)$ scaling corrections of the nonperturbative methods).
Many recipes of improvement
have been proposed (see e.g. \cite{L-M}, and \cite{C-L-V} for a review of them)
that essentially consist in a better choice of the expansion parameter.
Among them we mention the so-called tadpole improvement~\cite{L-M}
(MFI) motivated by mean-field arguments,
in which one scales the link variable with $u_0(g_0^2)\equiv 
\langle \case{1}{N} {\rm Tr}\, U_{x,\mu\nu}\rangle^{1/4}$ as measured
in the Monte Carlo simulation.
Accordingly one rescales
the coupling constant: $g_0^2\rightarrow g_{\rm mf}^2 = g_0^2/u_0^4$.
Thus, if at one loop: 
$Z = 1 + z_1 g_0^2 +O(g_0^4)$,
one obtains a mean-field improved expansion by 
\begin{equation}
Z =  u_0\left[ 1 + g_{\rm mf}^2 \left( z_1 + {1\over 12}\right)
+O(g_{\rm mf}^4)\right]
\end{equation}
For example, for $SU(3)$ in the quenched approximation and at $g_0^2=1$
one finds $u_0\simeq 0.878$ and  $g_{\rm mf}^2\simeq 1.68$.
A more naive and simple recipe of improvement consists just in 
the change of variable $g_0\rightarrow g_{\rm mf}$ in the standard perturbative expansion
(NMFI).

In the context of the clover action,
the following improved lattice operators have been considered~\cite{Heatlie-et-al}:
\begin{equation}
\psi \left[ 1 + {1\over 4} \left( \gamma_\alpha \stackrel{\leftarrow}{D}_\alpha - 
m_0\right)\right] 
\lambda^a \Gamma \left[ 1 - {1\over 4} \left( \gamma_\beta \stackrel{\rightarrow}{D}_\beta + 
m_0\right)\right]\psi 
\label{Lattop}
\end{equation}
where $\Gamma=\gamma_\mu,\gamma_\mu\gamma_5$ for $V_\mu^a$ and $A_\mu^a$ respectively,
and $D_\mu$ is the symmetric lattice covariant derivative.
Their one-loop renormalization is known~\cite{B-F-G-P}
\begin{equation}
Z_{V,A} = 1 + z_{V,A} g_0^2 + O(g_0^4),
\label{ttt}
\end{equation}
where
\begin{eqnarray}
&&z_{V}(c_{\rm{SW}}) = {c_F\over 16\pi^2} \left( -14.36 + 3.30\, c_{\rm{SW}}
  - 0.75\, c_{\rm{SW}}^2\right), \\
&&z_{A}(c_{\rm{SW}}) = {c_F\over 16\pi^2} \left( 6.87 - 12.54\, c_{\rm{SW}}  
+ 3.55\, c_{\rm{SW}}^2\right).
\label{onec}
\end{eqnarray}
The cactus dressing of the above one-loop expressions can be simply obtained by
using the dressed transverse gluon propagator (\ref{propdr})
and by rescaling $c_{\rm{SW}}$ according
to Eq.~(\ref{cswresc}). We thus obtain the following approximate 
expressions
\begin{equation}
Z_{V,A} \approx 1 + g_0^2 {z_{V,A}(\bar{c}_{\rm{SW}})\over 1-w(g_0^2)} 
\label{cadr}
\end{equation}

Nonperturbative  numerical calculations of $Z_{V,A}$  
for the tree-improved clover action (i.e.  $c_{\rm{SW}}=1$) and at $g_0^2=1$
have been obtained in quenched theory by imposing vector (V) and axial (A) WI's and by
nonperturbative renormalization on quark states (NP). 
In the Table we list these results and compare
them with the one-loop perturbative calculation (PT),
our cactus dressing (CI) of the one-loop expression, the 
mean-field inspired improvement (MFI)
and the result that one obtains just by substituting $g_0^2$ with $g_{\rm mf}^2$ (NMFI).
Results from other recipes can be found in Ref.~\cite{C-L-V}.
In the case of $Z_V$ all improved perturbative estimates get closer
to the nonperturbative results, thus improve PT.
On the other hand, in the case of $Z_A$ the simple change of
coupling from $g_0$ to $g_{\rm mf}$ (NMFI) does not help. 
Since $g_{\rm mf} > g_0$, it increases
the one-loop perturbative correction that has the ``wrong'' sign,
thus worsening the plain one-loop estimate.
Similarly, a change of coupling and momentum scale, in the manner of Lepage and
Mackenzie ~\cite{L-M}, also worsen the PT estimate as the corresponding
$g(q^*)$ (defined in \cite{L-M}) turns out to be larger than $g_0$.
In the case of $Z_A$ the only procedure improving PT is cactus resummation, 
but its estimate is still relatively far from the
nonperturbative result.

In Ref.~\cite{L-S-S-W} the lattice renormalizations of two further lattice operators
corresponding to $V_\mu$ and $A_\mu$ 
have been calculated nonperturbatively employing
finite size scaling techniques and using the nonperturbative estimate
of $c_{\rm SW}$.
The lattice operators were
\begin{eqnarray}
&&V_\mu^L = \bar{\psi}\lambda^a \gamma_\mu \psi + c_V {1\over 2} \left( \Delta^-_\mu + 
\Delta^+_\mu \right) i\bar{\psi}\lambda^a \sigma_{\mu\nu}\psi \\
&&A_\mu^L = \bar{\psi}\lambda^a \gamma_\mu \gamma_5\psi + c_A {1\over 2} \left( \Delta^-_\mu + 
\Delta^+_\mu \right) \bar{\psi}\lambda^a \gamma_5 \psi,
\end{eqnarray}
where $c_{V,A}$ are $O(g_0^2)$ constants, and the corresponding terms serve
to obtain on-shell improved operators.
Their perturbative renormalization is given by formula (\ref{ttt})
with~\cite{Cap-et-al-1}
\begin{eqnarray}
&&z_{V}(c_{\rm{SW}}) = {c_F\over 16\pi^2} \left( -20.62 + 
4.75\, c_{\rm{SW}}  + 0.54\, c_{\rm{SW}}^2\right), \\
&&z_{A}(c_{\rm{SW}}) = {c_F\over 16\pi^2} \left( -15.80 - 0.25 \, c_{\rm{SW}}  + 
2.25\, c_{\rm{SW}}^2\right).
\label{onec2}
\end{eqnarray}
In Fig.~2 we compare the nonperturbative calculations of Ref.~\cite{L-S-S-W}
with the one-loop and the dressed one-loop calculations. A remarkable improvement
is observed. In this case the cactus resummation performs
as the mean-field inspired boosted perturbation theory (MFI).
As already noted in Ref.~\cite{L-S-S-W}, the nonperturbative data are best
reproduced by NMFI.

It is clear that nonperturbative methods are in general 
preferable to approximations based on perturbative calculations, due
to their better controlled systematic errors
($O(a)$ against $O(g_0^n)$). However, improved perturbative estimates
are still quite
useful. They indeed provide important consistency checks.
Further, in those cases where nonperturbative methods
are difficult to implement, perturbative methods remain
the only source of quantitative information.
In this report we have shown how to extend to
the clover improved lattice formulation of QCD
the resummation of cactus diagrams, which represents
a direct implementation of the idea of tadpole dominance.
The examples considered here and in Ref.~\cite{P-V}
show that the resummation of cactus 
diagrams leads to a general improvement in the evaluation of the 
lattice renormalizations based on perturbation theory.
The comparison with the corresponding nonperturbative calculations
is globally satisfactory.
Of course, cactus resummation may also be applied to the
lattice  renormalizations of other operators without
further complications.

% ========================= REFERENCES =========================

\break

\centerline{FIGURES}
\medskip
\bigskip
\centerline{\psfig{figure=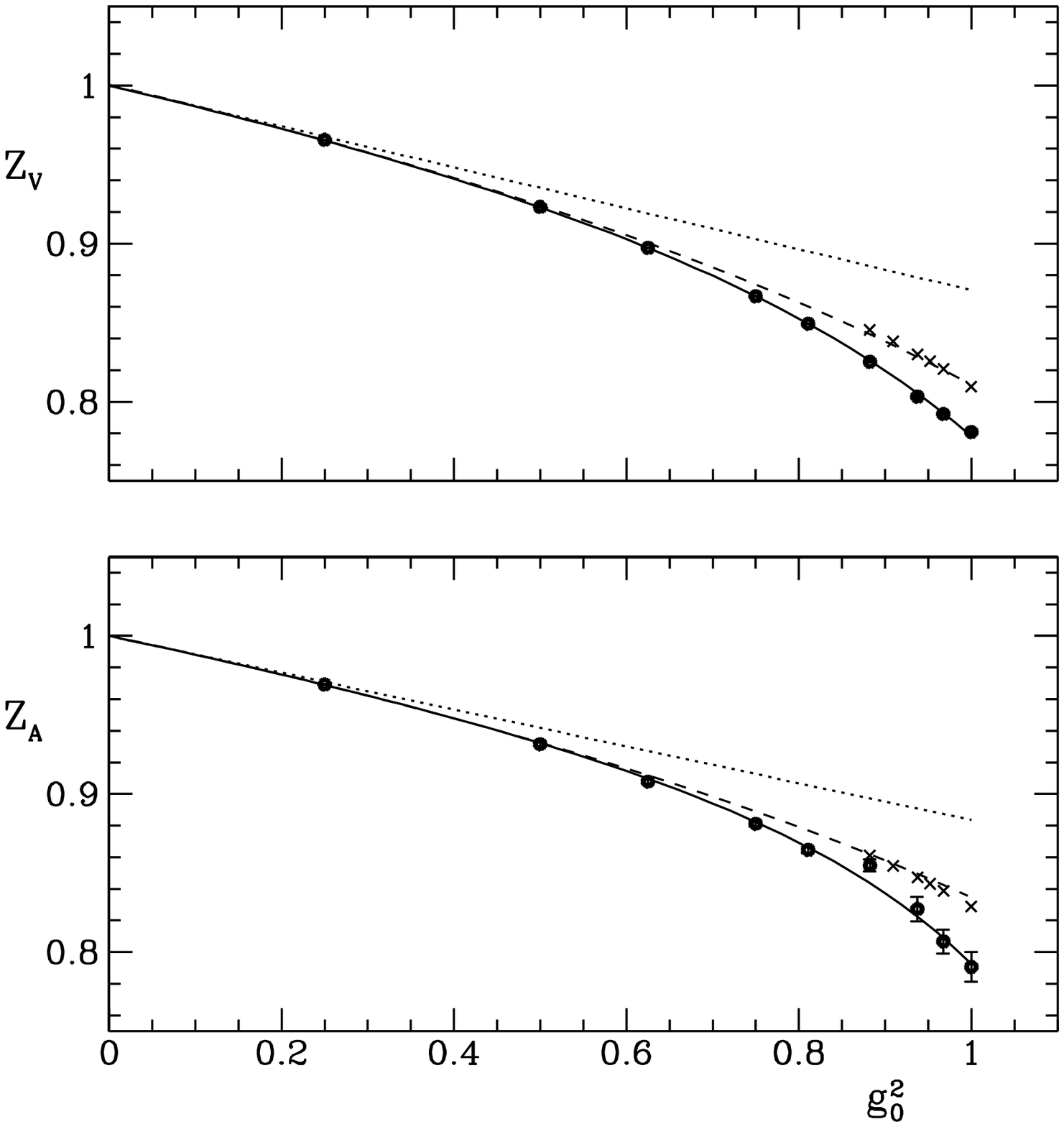,width=20truecm}}
\begin{enumerate}
\item[Figure 2: ] Results for $Z_V$ and $Z_A$ (from
Ref.~\cite{L-S-S-W}), coming from numerical simulations (filled
circles, fitted by a solid line), bare perturbation theory (dotted lines) and ``mean
field improved" perturbation theory (crosses). The dashed lines
superimposed on these figures are our results from cactus dressing.

\end{enumerate}

% ========================= TABLES =========================

\begin{table}
%\squeezetable
\caption{ Some estimates of $Z_V$ and $Z_A$ for the operators (\ref{Lattop})
and the tree-improved clover action at $g_0^2=1$ ($\beta=6$). 
\label{tab1}}
\begin{tabular}{lr@{}lr@{}l}
\multicolumn{1}{c}{Method}&
\multicolumn{2}{c}{$Z_V$}&
\multicolumn{2}{c}{$Z_A$}\\
\tableline \hline
PT & 0&.90 & 0&.98 \\
CI & 0&.86 & 1&.00 \\
MFI& 0&.85 & 0&.97 \\
NMFI& 0&.83 & 0&.97 \\
VWI \cite{C-L-V} & 0&.82 & & \\
AWI \cite{C-L-V} & 0&.80(2) & 1&.11(2) \\
NP \cite{M-P-S-T-V} & 0&.84(1) & 1&.06(8) 
\end{tabular}
\end{table}

\end{document}